\documentclass[aps,prl,showpacs,twocolumn]{revtex4}
\usepackage{graphicx}
\usepackage{dcolumn}
\usepackage{xspace}

\begin{document}

\title{Half-metallic ferromagnetism and structural stability of zincblende
phases of the transition-metal chalcogenides}
\author{Wen-Hui Xie}
\author{Ya-Qiong Xu}
\author{Bang-Gui Liu}
\email[Corresponding author. \\ Email address:
]{bgliu@aphy.iphy.ac.cn} \affiliation{ Institute of Physics \&
Center of Condensed Matter Physics, Chinese Academy of Sciences,
Beijing 100080, P. R. China}
\author{D. G. Pettifor}
\affiliation{ Department of Materials, University of Oxford, Parks
Road, Oxford OX1 3PH, UK}

\date{\today}
\widetext

\begin{abstract}
An accurate density-functional method is used to study
systematically half-metallic ferromagnetism and stability of
zincblende phases of 3d-transition-metal chalcogenides. The
zincblende CrTe,  CrSe, and VTe phases are found to be excellent
half-metallic ferromagnets with large half-metallic gaps (up to
0.88 eV). They are mechanically stable and approximately 0.31-0.53
eV per formula unit higher in total energy than the corresponding
nickel-arsenide ground-state phases, and therefore would be grown
epitaxially in the form of films and layers thick enough for
spintronic applications.
\end{abstract}

\pacs{75.90.+w, 62.25.+g, 73.22.-f, 75.30.-m \hspace{2.5cm} Phys
Rev Lett 91, 037204 (2003)} \maketitle

Half-metallic ferromagnets are seen as a key ingredient in future
high performance spintronic devices, because they have only one
electronic spin channel at the Fermi energy and, therefore, may
show nearly 100 \% spin polarization\cite{sptr,HMM}. Since de
Groot {\it et al}'s discovery\cite{heusler} in 1983, a lot of
half-metallic ferromagnets have been theoretically predicted and
some of them furthermore have been confirmed experimentally
\cite{NiMnSb,CrO2,Fe3O4,LSMO}. Much attention has been paid to
understanding the mechanism behind the half-metallic magnetism and
to studying its implication on various physical
properties\cite{phys,groot1}. However, it is highly desirable to
explore new half-metallic ferromagnetic materials which are
compatible with important III-V and II-VI semiconductors. For this
purpose, effort has be made on the metastable zincblende (B3)
phases such as the transition-metal pnictides
\cite{epi-mnx,mnasdot,nhmm,cras,cras1,crsb,crsb1,mnx,shirai,lbg,xu}.
Although zincblende phases of MnAs\cite{mnasdot},
CrAs\cite{cras,cras1} and CrSb\cite{crsb} have been successfully
fabricated as nanodots, ultrathin films and ultrathin layers in
multilayers, it has not been possible to grow the zincblende
half-metallic ferromagnetic phases as high-quality layers or thick
films. This is due to the metastable zincblende phases being about
1 eV per formula unit higher in energy than the ground state
nickel-arsenide (B8$_1$) phases. However, spintronic devices
require thick films or layers. Therefore, it is important to
explore theoretically other half-metallic ferromagnetic materials,
which on the one hand are compatible with the binary
tetrahedral-coordinated semiconductors, and on the other hand are
not only low in energy with respect to the corresponding
ground-state structures but also mechanically stable against
structural deformations.

In this Letter we make use of an accurate full-potential
density-functional method to study systematically transition-metal
chalcogenides in the zincblende and nickel-arsenide structures in
order to find half-metallic ferromagnetic phases which could be
realized in the form of films and layers thick enough. We shall
show that CrTe, CrSe, and VTe in the zincblende structure are
excellent half-metallic ferromagnets with wide half-metallic gaps.
They will be proved to be mechanically stable and approximately
0.31-0.53 eV per formula unit higher in energy than the
corresponding ground-state phases, and therefore would be grown
epitaxially in the form of films and layers thick enough for
spintronic applications.

We make use of the Vienna package WIEN2k\cite{wien2k} for all our
calculations. This is a full-potential (linear) augmented plane
wave plus local orbitals method within the density functional
theory\cite{DFT}. We take the generalized gradient approximation
in Ref \cite{PBE96} for the exchange-correlation potential.
Relativistic effects are taken into account within the scalar
approximation, but the spin-orbit coupling is neglected because it
is proved to have little effect on our main conclusions. We use
3000 k points in the Brillouin zone for the zincblende structure
and 2000 k points for the nickel-arsenide structure. When
calculating the shear modulus constants of the zincblende phases
we use 6000 k points. We set $R_{mt}*K_{max}$ to 8.0 and make the
expansion up to $l=10$ in the muffin tins. The self-consistent
calculations are considered to be converged only when the
integrated charge difference per formula unit,
${\int{}|\rho_n-\rho_{n-1}|dr}$, between input charge density
(${\rho_{n-1}(r)}$) and output (${\rho_{n}(r)}$) is less than
0.0001.

In order to search for the better half-metallic ferromagnets in
zincblende structure, we explore systematically all zincblende
phases of 3d-transition-metal chalcogenides. We find that among
all these zincblende compounds only the CrSe, CrTe and VTe phases
are half-metallic ferromagnets. The ground-state phase of CrTe is
a metallic ferromagnet in the hexagonal nickel-arsenide structure
with experimental lattice constants a=3.998 {\AA} and c=6.254
{\AA}\cite{crtesee,crteset}. Its experimental Curie temperature is
Tc=340 K which decreases to zero at a pressure of 28
kbar\cite{crtesee}. In contrast, the ground-state phase of CrSe is
an antiferromagnet in the nickel-arsenide structure with
experimental lattice constants a=3.674 {\AA} and c=6.001
{\AA}\cite{crse,crtesee}. Its Neel temperature was located at 320
K by specific heat measurement\cite{crseneel}. There has been not
any experimental report on VTe, but it is shown by comparing the
total energies of various phases that the nickel-arsenide
ferromagnetic phase, with equilibrium lattice constants 4.13 and
6.07 \AA, is the ground-state phase in this case. We predict the
equilibrium lattice constants of the zincblende CrSe, CrTe, and
VTe to be 5.833, 6.292, and 6.271 {\AA}, respectively.

\begin{figure}[tp]
\includegraphics[width=8cm]{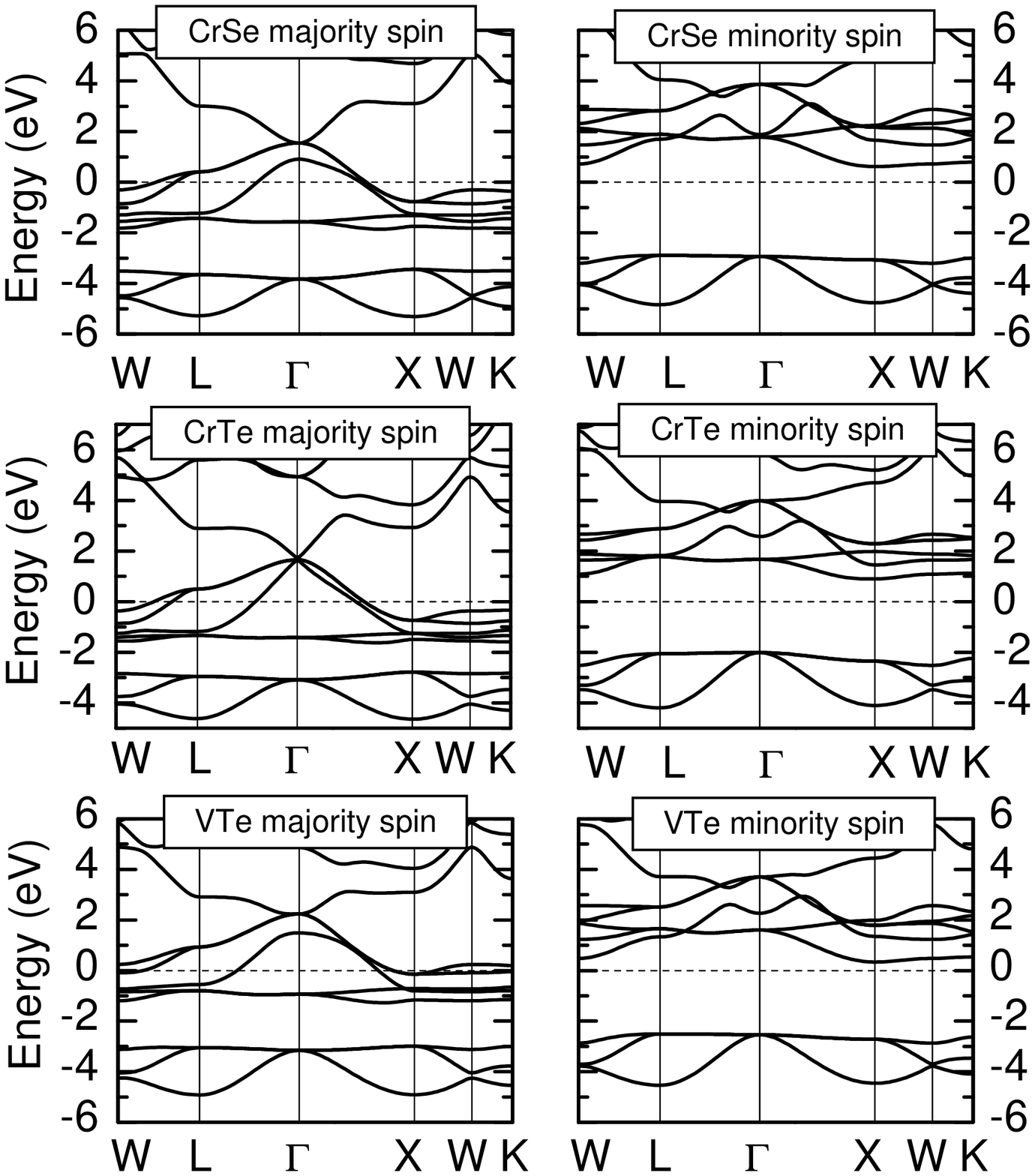}
\caption{The band structures of the zincblende phases of CrSe
(upper panels), CrTe (middle panels) and VTe (lower panels) at
their equilibrium lattice constants. The left panels are the
majority-spin bands and the right panels the minority-spin bands.}
\label{fig01}
\end{figure}
Figure 1 shows the band structures of the zincblende CrSe, CrTe,
and VTe phases at their equilibrium volumes. For the minority spin
bands we see that just below the Fermi energy there exist three
$\Gamma_{15}$ bands which result mainly from the Te (Se) $p$
electrons, whereas just above the Fermi energy there exist two
$\Gamma_{12}$ bands which comprise mainly the Cr (V) $e_g$
electrons. For the majority-spin bands, there are also three
$\Gamma_{15}$ bands below the Fermi energy originating mainly from
the Te (Se) $p$ electrons, but the two $\Gamma_{12}$ bands mainly
of the Cr (V) $e_g$ electrons are below the Fermi energy. The
bands crossing the Fermi energy are the majority-spin $\Gamma_{1}$
and $\Gamma_{15}'$. The three $\Gamma_{15}'$  bands are mainly of
Cr (V) $t_{2g}$ character. There is strong interaction between the
Te (Se) $p$ bands and the Cr (V) $t_{2g}$ bands. Figure 2 shows
the spin-dependent total and partial DOS of the three zincblende
phases. The zincblende phases of CrTe, CrSe, and VTe have large
half-metallic gaps, 0.88, 0.61, and 0.31 eV, respectively.

\begin{figure}[bp]
\includegraphics[width=8cm]{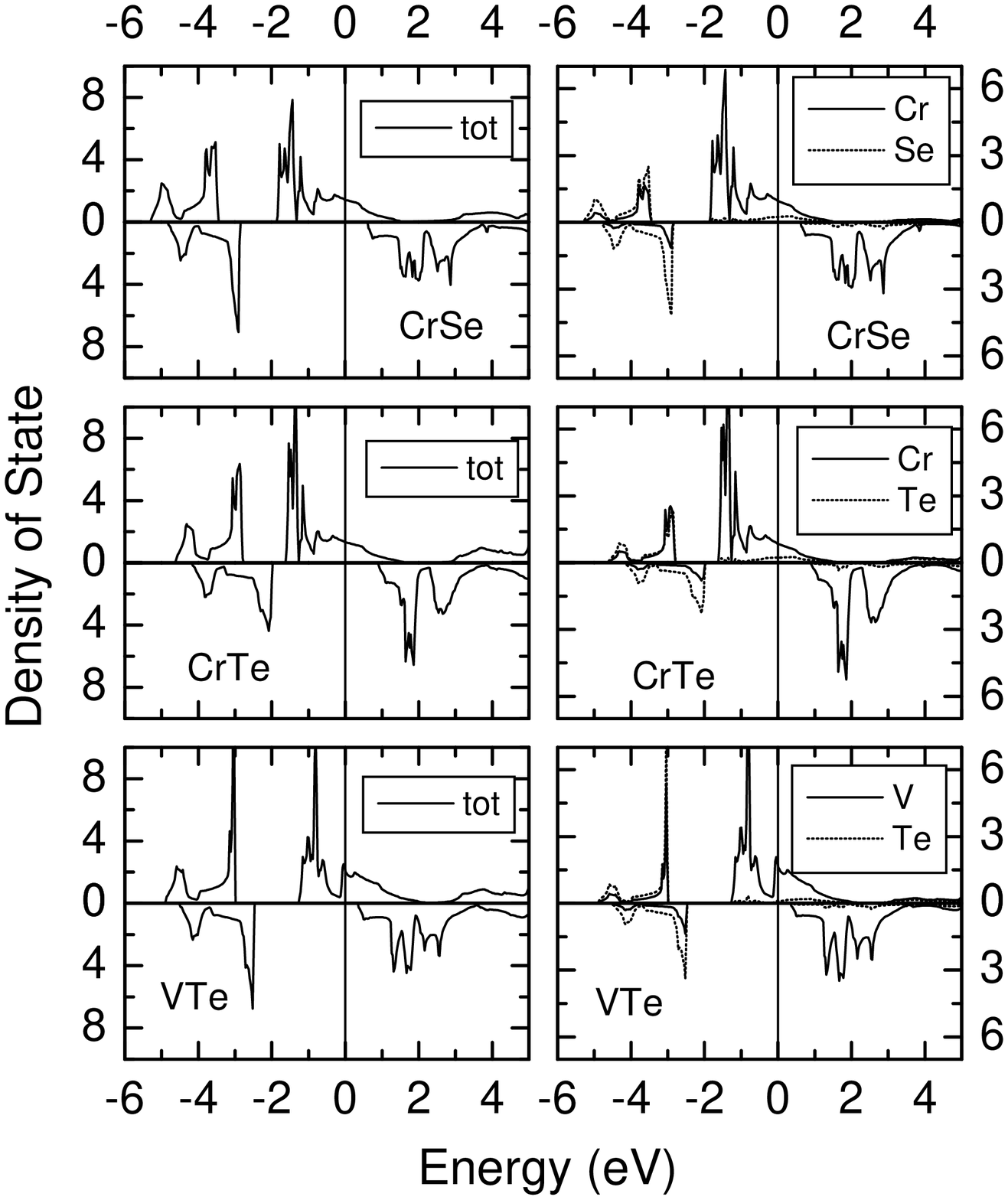}
\caption{Spin-dependent total (tot) and partial DOS (per eV per
formula unit) of CrSe, CrTe, and VTe in the zincblende structure.
The upper-left panel shows the total DOS of the CrSe phase, and
the upper-right panel the Cr (solid lines) and Se (dot lines)
partial DOS; the middle-left panel shows the total DOS of the CrTe
phase, and the middle-right panel the Cr (solid lines) and Te (dot
lines) partial DOS; the lower-left panel shows the total DOS of
the VTe phase, and the lower-right panel the V (solid lines) and
Te (dot lines) partial DOS.} \label{fig02}
\end{figure}
The binding energy curves for CrSe, CrTe, and VTe in the
zincblende structure are shown in Figure 3. All the three
zincblende phases are ferromagnetic because the antiferromagnetism
with the modulation vector [001] makes their equilibrium total
energy per formula unit increase by 0.07, 0.12, or 0.13 eV,
respectively, and other antiferromagnetic modulation vectors even
lead to higher total energies. We summarize our main results in
Table I. The metastable energy of a phase is defined as its total
energy per formula unit minus that of the corresponding
ground-state phases. There have been previous predictions of
half-metallic ferromagnetism in transition-metal pnictides or
chalcogenides from DOS calculations\cite{crte3}. However, no
detailed discussion of their energetics has been given. It is
clear that the metastable energies of the three transition-metal
chalcogenides are substantially smaller than those  of the
transition-metal pnictides.

\begin{figure}[tp]
\includegraphics[width=8cm]{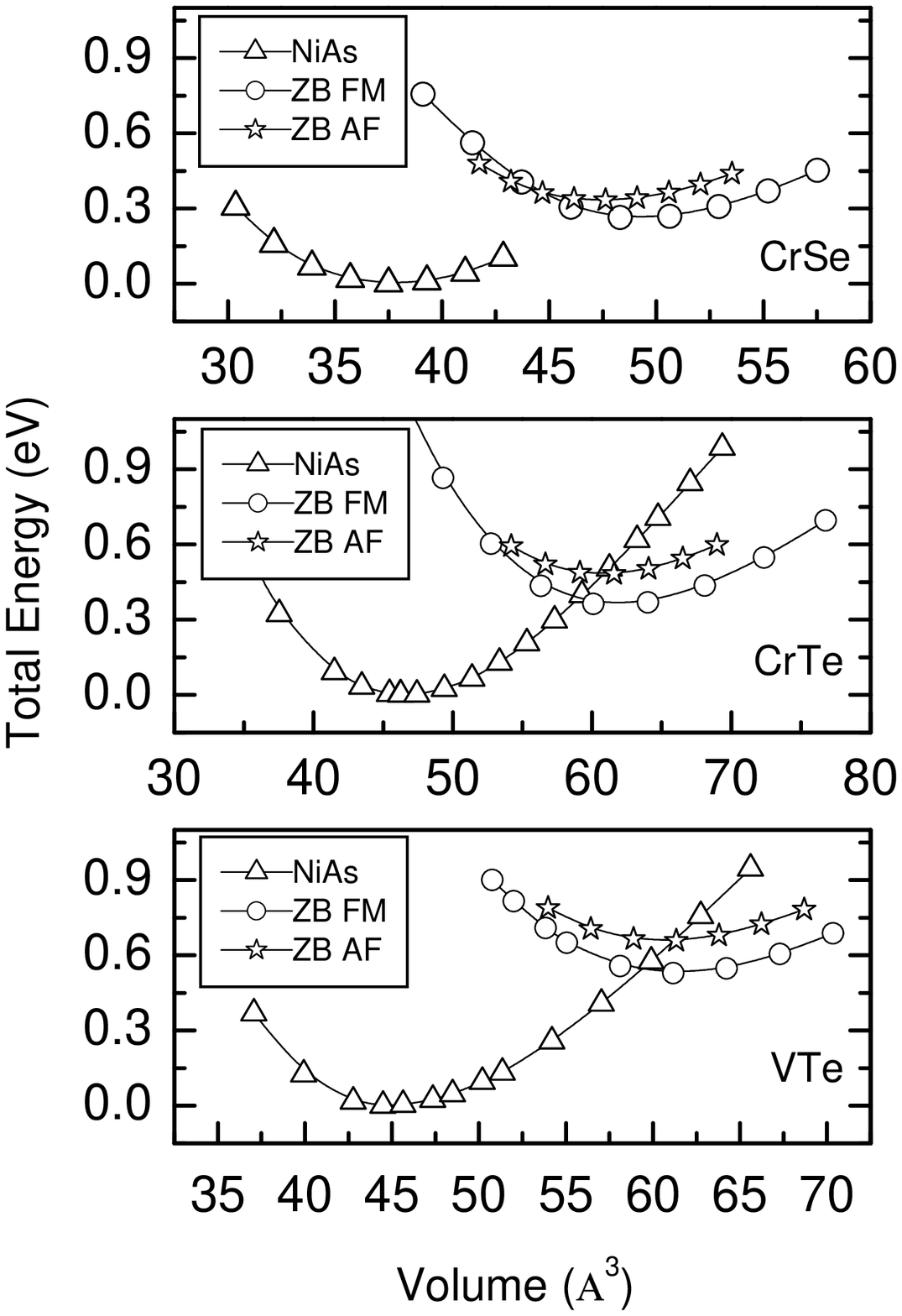}
\caption{The total energies of the zincblende (ZB) phases, with
respect to the corresponding NiAs (triangles) ground-state phases,
as functions of the volume per formula unit for CrSe (upper
panel), CrTe (middle panel), and VTe (lower panel). The
half-metallic ferromagnetism in the ZB ferromagnetic phases (ZB
FM, circles) persists to be nonzero up to a compression of 12 \%,
22 \%, or 10\% about the equilibrium volume, respectively. The ZB
antiferromagnetic phases (ZB AF, stars), which are lowest in
energy among all the antiferromagnetic ZB structures we can
construct, are presented for comparison. } \label{fig03}
\end{figure}

\begin{table}[tbp]
\caption{\label{tab:table1} The predicted equilibrium lattice
constant ($a$), magnetic moment per formula unit ($M$),
half-metallic gap ($E_g$), and metastable energy ($E_t$) of the
transition-metal pnictides and chalcogenides with zincblende
structure. }
\begin{ruledtabular}
\begin{tabular}{ccccc}
 name  &  $a$ (\AA) & $M$ ($\mu_B$) & $E_g$ (eV) & $E_t$ (eV) \\
\hline MnAs\cite{nhmm} &  5.70   & 3.5    & -    & 0.9   \\
\hline MnSb\cite{xu} &  6.18   & 4.000    & 0.20    & 0.9   \\
\hline MnBi\cite{xu} &  6.399   & 4.000    & 0.42    & 1.0  \\
\hline CrAs &  5.659   & 3.000\cite{shirai}    & 0.46    & 0.93   \\
\hline CrSb\cite{lbg} &  6.138   & 3.000    & 0.77    & 1.0   \\
\hline VTe  &  6.271   & 3.000    & 0.31    & 0.53   \\
\hline CrSe &  5.833   & 4.000    & 0.61    & 0.31   \\
\hline CrTe &  6.292   & 4.000    & 0.88    & 0.36
\end{tabular}
\end{ruledtabular}
\end{table}

An important question to address is whether these zincblende
phases are mechanically stable, so that they would be grown
experimentally\cite{shear}. In order to investigate this we have
evaluated the tetragonal and trigonal shear constants, $C'$ and
C$_{44}$, respectively, by computing the change in energy of the
zincblende phases under small volume-conserving strains (see
Figure 4). Their values are given in table II together with the
bulk modulus $B$ for comparison. We see that all the zincblende
phases are stable against the tetrahedral and rhombohedral
deformations. The three zincblende transition-metal chalcogenides
have slightly smaller bulk moduli, but larger tetragonal shear
moduli than the zincblende CrAs phase. They are softer than GaAs,
which has bulk and tetragonal shear modulus of 61.3 and 59.7 GPa,
but are harder than the zincblende CrAs phase which has already
been fabricated successfully epitaxially. Importantly, however,
their metastable energies are much smaller than that of the
zincblende CrAs phase. Therefore, the zincblende phases of CrSe,
CrTe and VTe would be realized  in the near future by means of
epitaxial growth.

\begin{figure}[bp]
\includegraphics[width=8cm]{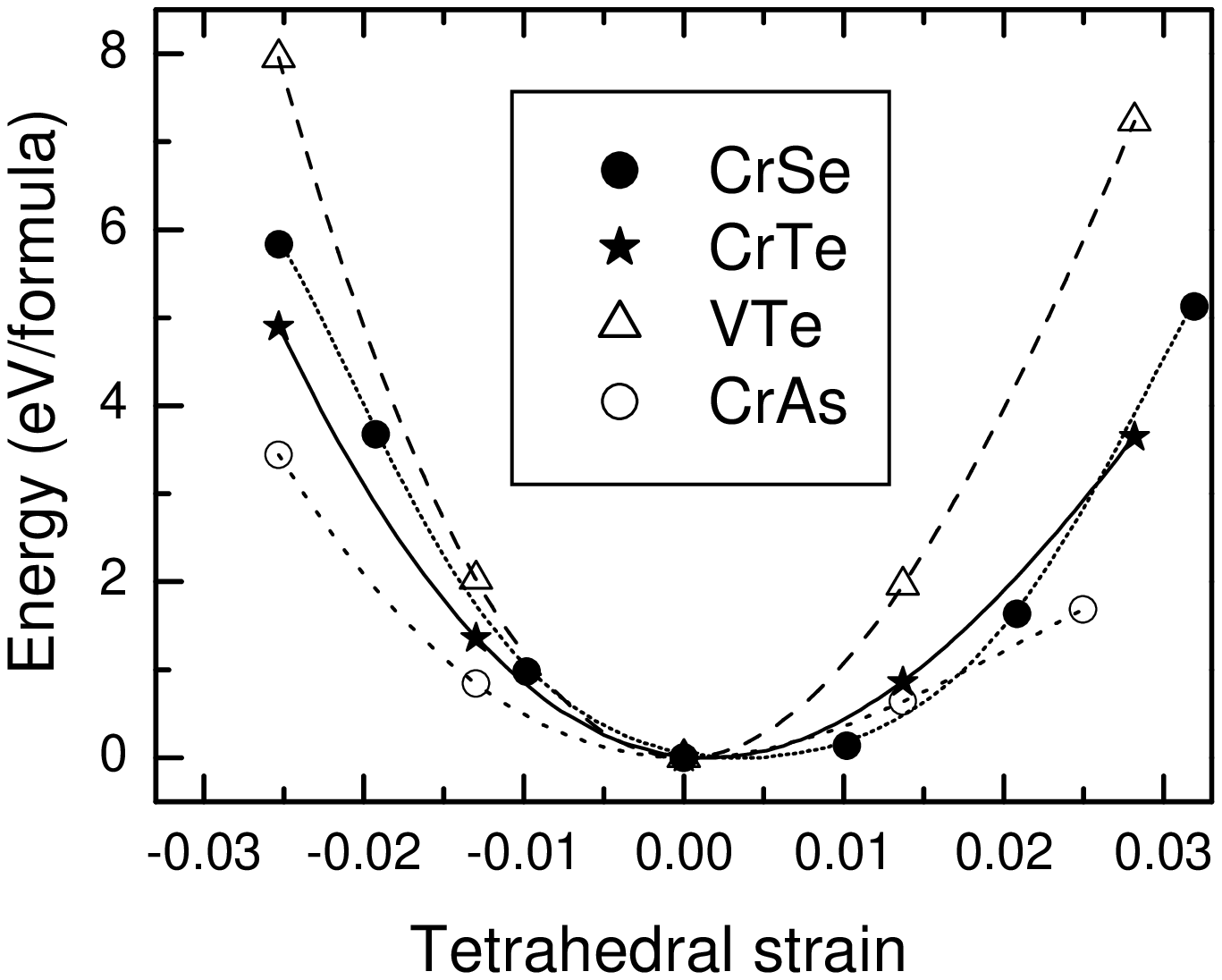}
\caption{Total energies of the equilibrium zincblende phases of
CrSe (dot line plus solid circles), CrTe (solid line plus stars),
VTe (long dash plus triangles) and CrAs (dash line plus open
circles) as functions of tetrahedral strain at fixed atomic
volume.} \label{fig04}
\end{figure}

\begin{table}[tb]
\caption{\label{tab:table2} The predicted bulk modulus ($B$) and
shear moduli ($C'$ and $C_{44}$) of the zincblende
transition-metal chalcogenides with half-metallic ferromagnetism.
CrAs is presented for comparison.}
\begin{ruledtabular}
\begin{tabular}{cccc}
 name      &   $B$ (GPa)  &    $C'$ (GPa) &  $C_{44}$ (GPa)  \\
\hline VTe &    50.3    &       9.9    &       30.5  \\
\hline CrSe &   59.5      &      5.6      &     50.7 \\
\hline CrTe &   45.9      &      5.5       &    36.4 \\
\hline CrAs &   71.0      &      5.1      &     46.1
\end{tabular}
\end{ruledtabular}
\end{table}

In summary, using the accurate density-functional method we have
made a systematic computation of all the 3d-transition-metal
chalcogenides in the zincblende and nickel-arsenide structures and
found the zincblende CrSe, CrTe and VTe to be excellent
half-metallic ferromagnets. These zincblende phases have very
large half-metallic gaps (up to 0.88 eV) and  are approximately
0.31-0.53 eV per formula unit higher in energy than the
corresponding nickel-arsenide ground-state phases. Moreover, they
are mechanically stable and harder than the observed zincblende
CrAs phase. Therefore, they would be grown epitaxially in the form
of films and layers thick enough, and useful in spintronics and
other applications.

\begin{acknowledgments}
This work is supported in parts by Chinese Department of Science
and Technology under the National Key Projects of Basic Research
(No. G1999064509), by Nature Science Foundation of China (No.
60021403), and by British Royal Society under a collaborating
project with Chinese Academy of Sciences.
\end{acknowledgments}

\end{document}